\documentclass{sigchi-ext}
\usepackage[T1]{fontenc}
\usepackage{textcomp}
\usepackage[scaled=.92]{helvet} 
\usepackage{graphicx} 
\usepackage{balance}  
\usepackage{booktabs} 
\usepackage{ccicons}  
\usepackage{ragged2e} 

\copyrightinfo{\scriptsize Permission to make digital or hard copies of part or all of this work for personal or classroom use is granted without fee provided that copies are not made or distributed for profit or commercial advantage and that copies bear this notice and the full citation on the first page. Copyrights for third-party components of this work must be honored. For all other uses, contact the owner/author(s). \\
{\emph{CHI '20 Extended Abstracts, April 25--30, 2020, Honolulu, HI, USA.}} \\
\copyright~2020 Copyright is held by the author/owner(s). \\ 
ACM ISBN 978-1-4503-6819-3/20/04. \\
http://dx.doi.org/10.1145/3334480.3382980}

\def\plaintitle{Are All the Frames Equally Important?} 
\def\emptyauthor{}
\def\plainkeywords{Attention; video; saliency; temporal saliency; eye-tracking}

\title{\plaintitle}

\numberofauthors{6}
\author{%
  \alignauthor{%
    \textbf{Oleksii Sidorov}\\
    \textbf{Marius Pedersen}\\
    \affaddr{The Norwegian Colour and Visual Computing Laboratory, NTNU} \\
    \affaddr{Gj{\o}vik, Norway} \\
    \email{oleksiis@stud.ntnu.no}
    \email{marius.pedersen@ntnu.no}}
    \alignauthor{%
    \textbf{Nam Wook Kim}\\
    \affaddr{Harvard University}\\
    \affaddr{Cambridge, USA}\\
    \email{namwkim@seas.harvard.edu} } \vfil
    \alignauthor{%
    \textbf{Sumit Shekhar}\\
    \affaddr{Adobe Research}\\
    \affaddr{San Jose, USA}\\
    \email{sushekha@adobe.com} } }

\definecolor{linkColor}{RGB}{6,125,233}
\hypersetup{%
  pdftitle={\plaintitle},
  pdfauthor={\emptyauthor},
  pdfkeywords={\plainkeywords},
  bookmarksnumbered,
  pdfstartview={FitH},
  colorlinks,
  citecolor=black,
  filecolor=black,
  linkcolor=black,
  urlcolor=linkColor,
  breaklinks=true,
}

\begin{document}


\maketitle

\RaggedRight{} 

\begin{abstract}
In this work, we address the problem of measuring and predicting temporal video saliency - a metric which defines the importance of a video frame for human attention. Unlike the conventional spatial saliency which defines the location of the salient regions within a frame (as it is done for still images), temporal saliency considers importance of a frame as a whole and may not exist apart from context. \\
The proposed interface is an interactive cursor-based algorithm for collecting experimental data about temporal saliency. We collect the first human responses and perform their analysis. As a result, we show that qualitatively, the produced scores have very explicit meaning of the semantic changes in a frame, while quantitatively being highly correlated between all the observers. \\
Apart from that, we show that the proposed tool can simultaneously collect fixations similar to the ones produced by eye-tracker in a more affordable way. Further, this approach may be used for creation of first temporal saliency datasets which will allow training computational predictive algorithms. The proposed interface does not rely on any special equipment, which allows to run it remotely and cover a wide audience.
\end{abstract}




\section{Introduction}
\begin{marginfigure}[208 pt]
  \begin{minipage}{\marginparwidth}
    \centering
    \includegraphics[width=\marginparwidth]{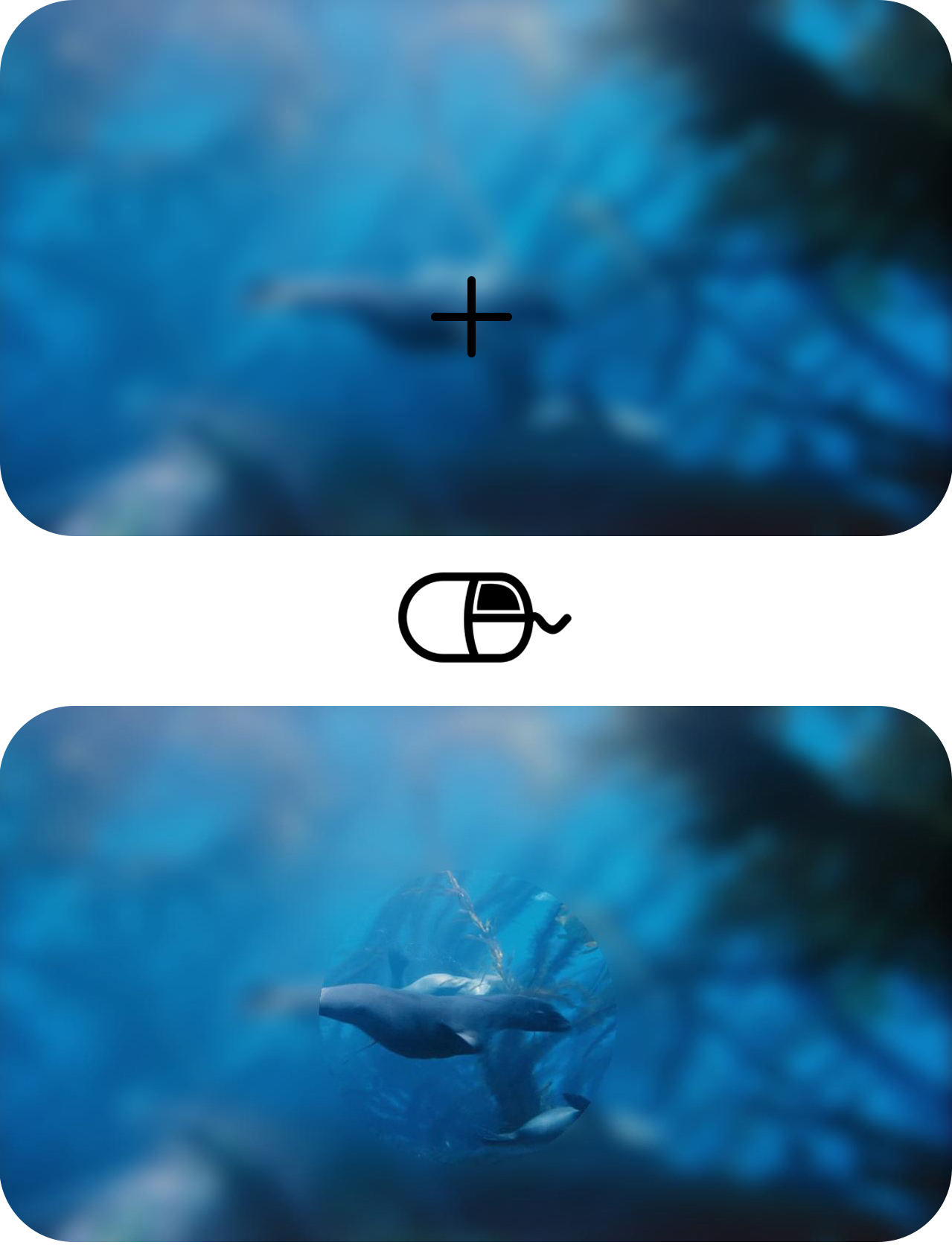}
    \caption{The proposed interface. A more representative video demonstration is available online: \href{https://drive.google.com/open?id=1IxoZ69ImmeguHQ5LWRroiQ93zyy6gqy3}{[link]}.}
      \label{fig:interface}
  \end{minipage}
\end{marginfigure}

It seems obvious that some fragments of a video are more important than others. Such fragments concentrate most of the viewer's attention while others remain of no interest. The na\"{i}ve examples are: a culmination scene in a movie, a screamer in a horror film, the moment of an explosion, or even a slight motion in very calm footage. We denote such fragments as groups of frames with high \emph{temporal saliency}. Information about temporal saliency is an essential part of a video characterization which gives valuable insights about the video structure. Such information is directly applicable in video compression (frames which do not attract attention may be compressed more), video summarization (salient frames contain the most of perceived video content), indexing, memorability prediction, and others tasks. So, the reader may expect that there is a big number of algorithms and techniques aimed at measuring and predicting temporal saliency. However, this is not the case. The most, if not all, of the well-known works on video saliency are aimed at spatial saliency, \textit{i.e.}, a prediction of spatial distribution of the observer's attention across the frame (in a similar way as if it was an individual image). We hypothesize that this is due to the absence of established methodology for measuring temporal saliency in the experiment, which is crucial for obtaining ground truth data. Conventionally, saliency data are collected using eye-tracking, which is a technique that produces a continuous temporal signal. In other words, it does not allow to differentiate between the frames as a whole, because each frame produces the same kind of output -- a pair of gaze fixation coordinates with a rate defined by hardware. \\
In this work, we propose a new methodology for measuring temporal video saliency in the experiment -- the first, to the best of our knowledge, method of this kind. For this, we develop a special interface based on mouse-contingent moving-window approach for measuring saliency maps of static images. We also show that it can simultaneously gather meaningful spatial information which can serve as an approximation of gaze fixations.\\
During the experiment, observers are presented with repeated blurry video-sequences which they can partially deblur using mouse click (Fig. \ref{fig:interface}). Users can deblur a circular region with a center at cursor location which approximates the confined area of focus in the human eye fovea surrounded by a blurred periphery \cite{gosselin2001bubbles}. Since the number of clicks is limited - observers are forced to use clicks only on most "interesting" frames which attract their attention. Statistical analysis of the collected clicks allows to assign the corresponding level of importance to each frame. This information can be applied directly in numerous tasks of video processing. \\
To summarize, unlike the conventional approaches which only try to understand \emph{where} the observer looks, we also study \emph{when} the observer pay the most attention.

\section{Related works}
The straightforward method of retrieving the information about attention is based on the utilization of commercial eye-trackers (\textit{e.g.} EyeLink, Tobii). Hardware-based eye-tracking has been used widely in various studies on human-computer interaction \cite{Jacob2003-JACETI}\cite{nielsen2010eyetracking}. A less accurate, but much more affordable, way of measuring saliency is based on measuring the mouse cursor position which was proven to correlate strongly with gaze fixations \cite{guo2010towards}\cite{huang2012user}\cite{Rodden:2008:ECP:1358628.1358797}. The most successful algorithms of this type utilize a moving-window paradigm, which masks information outside of the area adjacent to the cursor and requires a user to move the cursor (followed by a window around it) to make other regions visible. Such algorithms include Restricted Focus Viewer software by Jansen \textit{et al.} \cite{jansen2003tool} and more recent SALICON \cite{jiang2015salicon} and BubbleView \cite{kim2017bubbleview}. These algorithms were also used in large online crowdsourcing experiments due to the native scalability of cursor-based approaches. However, they were studied only in the context of spatial saliency of static images. This is fair for static images, but for video-sequences, temporal information is commonly even more important than spatial regions. Furthermore, there are no well-known experimental datasets which can provide this kind of information\footnote{A comprehensive list of saliency datasets: \href{http://saliency.mit.edu/datasets.html}{http://saliency.mit.edu/datasets.html}} and be used for training of computational algorithms. For example, the popular video saliency datasets Hollywood-2 \cite{6909754}, UCF sports \cite{mathe2015actions}, SAVAM \cite{Gitm1410:Semiautomatic}, DHF1K \cite{wang2018revisiting} only provide eye-tracking results which are constant in the temporal domain.

\section{Methodology}
Our approach is inspired by moving-window gaze approximations methods for still images. In the proposed setup all video frames are blurred. Clicking the mouse deblurs a round window around the cursor. Users are demonstrated repeated video sequence during which they can click the mouse for short periods of time. The total number of times when the frame was deblurred defines temporal saliency score, while location of the cursor when the mouse button is pressed approximates gaze fixation location and allows to detect what caused the interest.

\subsection{Discretization}
Short fragment of a video is more likely to attract user's attention rather than a single frame, so we let the users keep the mouse button pressed instead of clicking on each frame they find interesting. However, when not forced explicitly, observers tend to keep the mouse button pressed all the time, which is natural. Thus, to obtain variation of scores, it is crucial to restrict users artificially. Our solution is to simply limit the amount of deblurred frames (time period), after which clicking the mouse button stops working, and additionally limit the amount of deblurred frames per one continuous click. The users cannot see the limits, instead, they learn them during a test trial and then follow them intuitively.  For example, a 10-second video may have up to 4 seconds of deblurred frames, but no more than 1 second at once. In the result, a user can make 4 long clicks 1 second each or a larger number of short clicks, while we are guaranteed to have at least four discrete responses after one run.

\subsection{Repetition}
The idea of repeating the videos may be used to gather more responses from one observer and have richer statistics. Moreover, if a salient event happens at the end, the observer may reach the limit before seeing it, so it is necessary to make a second round. Also, eye-motion and cognitive processing are faster than clicking the mouse, so giving the user an opportunity to predict when an event will happen is beneficial for the creation of more accurate saliency maps with a shorter delay. However, we observed that in the majority of the cases, the first run is the most informative one, and the user is able to detect most salient information without preparation. \emph{Subsequent repeats lead to shifting the user's attention to smaller details.} Eventually, we used repetition in our experiments, but analyze different numbers of repeats in results.

\subsection{Other parameters}
Other important parameters are the blur radius and the radius of the window. Their definition requires more detailed study. The task given to an observer also influences where they look \cite{yarbus1967eye}\cite{kim2017bubbleview}, so, this parameter depends on the particular context in which the experiment is performed. In our case, we are interested in basic watching of a video without a particular task, so we worked under a "free-view" setup.

\begin{marginfigure}[62 pt]
  \begin{minipage}{\marginparwidth}
    \centering
    \includegraphics[width=\marginparwidth]{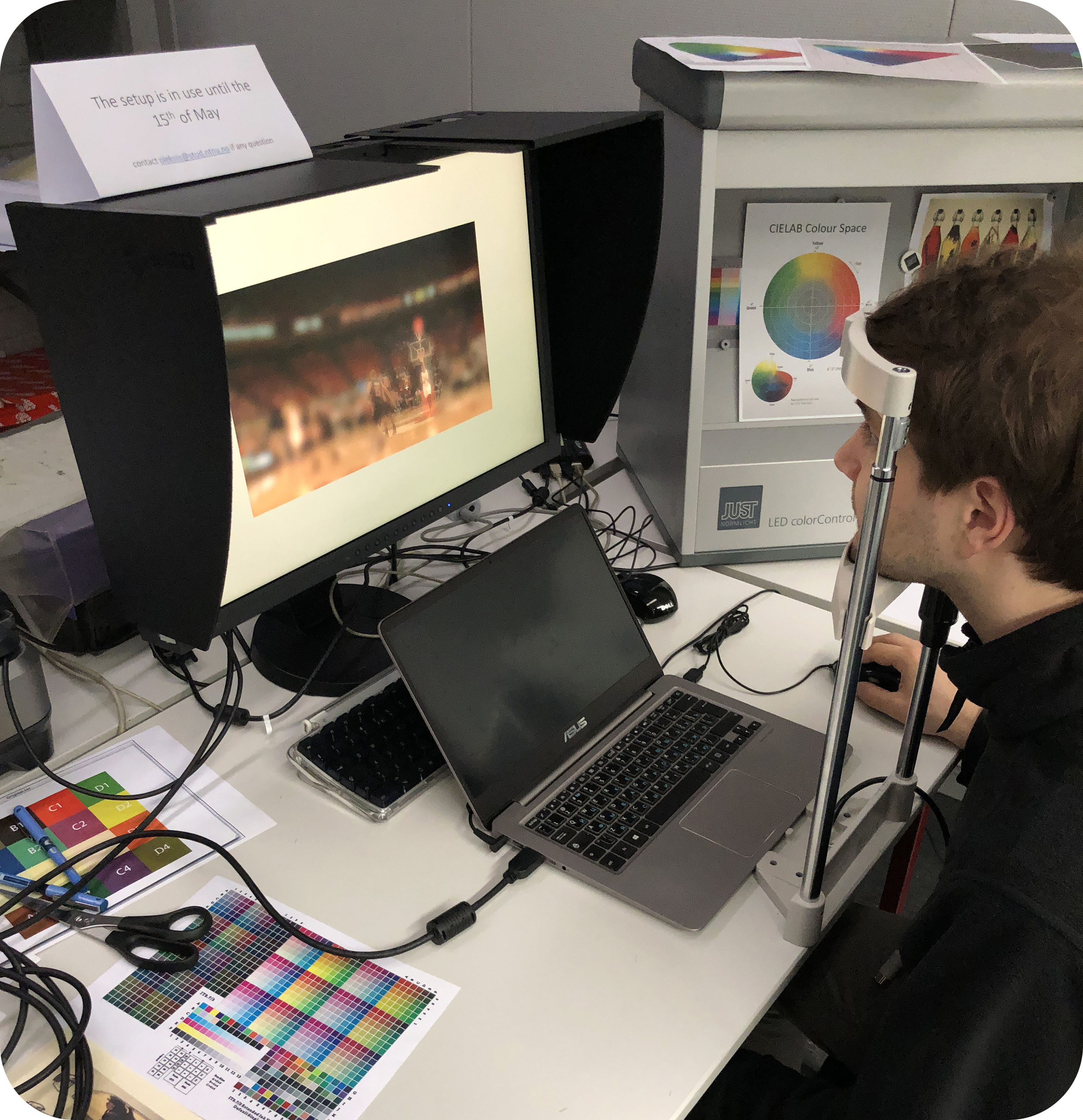}
    \caption{Experimental setup (the light is off during the session).}
      \label{fig:Tim}
  \end{minipage}
\end{marginfigure}

\section{Experimental setup}
The experiments were performed offline using a special setup in the laboratory (Fig. \ref{fig:Tim}) for the sake of fully-controlled conditions (in future we are also planning to run the experiment on Amazon Mechanical Turk for gathering larger database, which would be impossible to do with an eye-tracker). The display used is 24.1" EIZO ColorEdge CG241W color-calibrated with X-Rite Eye-One Pro. The distance between the display and the observer was 50 cm. \\
The code is written in MatLab with Psychtoolbox-3 \cite{kleiner2007s} and is publicly available by the link~\footnote{\href{https://github.com/acecreamu/temporal-saliency}{https://github.com/acecreamu/temporal-saliency}}.\\
Videos with ground-truth eye-tracking data were taken from SAVAM dataset \cite{Gitm1410:Semiautomatic} due to their high quality and diverse content. We used eight 10-seconds long HD videos including two test videos. The content of the videos is diverse and includes: a basketball game with a score moment, a calm shot of leaves in the wind, marine animals underwater, a cinematic scene of a child coming home, a surveillance camera footage of two men meeting, a suffocating diver emerging from the water. \\
Interface parameters: radius of a circular window -- 200 px ($6.2^{\circ}$ visual angle), blur kernel -- Gaussian with standard deviation of 15, video duration -- 10 s, limit of deblurred frames per one round -- 4 s (100 frames), limit of deblurred frames at one click -- 1 s (25 frames), number of repetitions -- 5, frame-rate of the videos -- 25 fps, video resolution -- 1280 px $\times$ 720 px ($38.2^{\circ} \times 22^{\circ}$ visual angle), videos are silent. \\
The observers were invited from the University staff and students. 30 subjects in total, 15 women and 15 men. Age: 21-42 (mean 25.6).

\begin{figure}
  \centering
  \includegraphics[width=0.99\columnwidth]{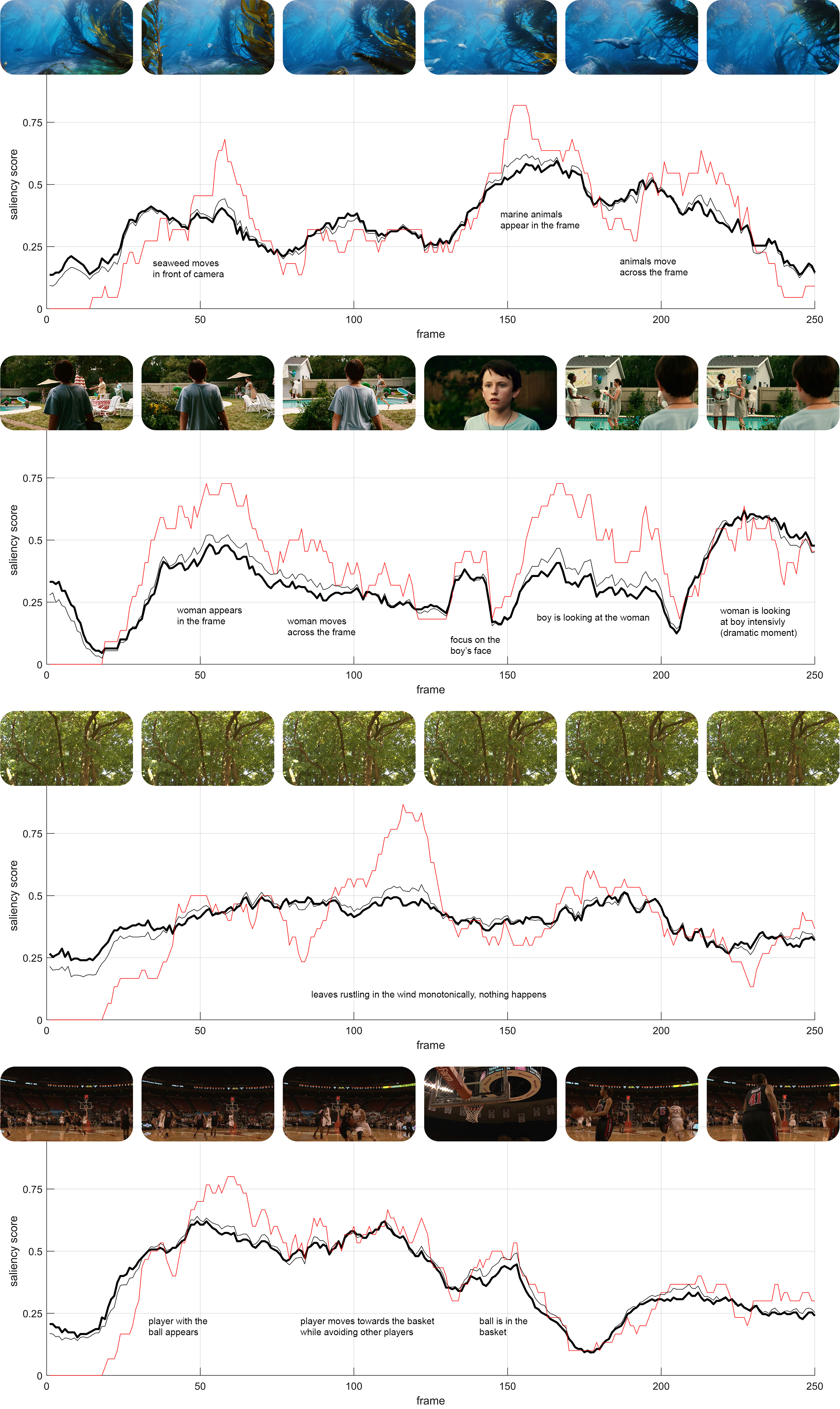}
  \caption{The produced temporal saliency graphs. Thick black line $C_{1-5}$, red line $C_1$, thin black line $C_{1-5}^{(W)}$. Zoom is required.}
  \label{fig:temporal}
\end{figure}

\begin{table*}[t!]
\small
  \begin{tabular}{p{3.8cm}rrrrp{0.4cm}rrrr}
    \toprule
    &  \multicolumn{4}{c}{\small{Pearson Correlation Coefficient (mean $_{std}$)}} & & \multicolumn{4}{c}{\small{Kolmogorov-Smirnov test (mean $p$-value)}}\\
     & $C_1$ & $C_{1-2}$ & $C_{1-5}$ & $C_{1-5}^{(W)}$ & & $C_1$ & $C_{1-2}$ & $C_{1-5}$  & $C_{1-5}^{(W)}$\\
    \midrule
    "The underwater world" & \textbf{0.663} $_{0.082}$ & \textbf{0.694} $_{0.082}$ & \textbf{0.740} $_{0.074}$ & \textbf{0.770} $_{0.064}$ & & 0.119 & 0.048 & 0.011 & 0.036\\
    "Cinematic scene" & \textbf{0.615} $_{0.092}$ & \textbf{0.711} $_{0.057}$ & \textbf{0.803} $_{0.053}$ & \textbf{0.789} $_{0.051}$ & & 0.164 & 0.107 & 0.033 & 0.067\\
    "Leaves in the wind" & \textbf{0.694} $_{0.068}$ & \textbf{0.563} $_{0.099}$ & \textbf{0.545} $_{0.108}$ & \textbf{0.647} $_{0.092}$ & & 0.081 & 0.073 & 0.044 & 0.057\\
    "Basketball game" & \textbf{0.741} $_{0.072}$ & \textbf{0.766} $_{0.070}$ & \textbf{0.863} $_{0.050}$ & \textbf{0.845} $_{0.051}$ & & 0.164 & 0.099 & 0.055 & 0.063\\
    "Diver suffocating" & \textbf{0.789} $_{0.050}$ & \textbf{0.788} $_{0.054}$ & \textbf{0.820} $_{0.057}$ & \textbf{0.834} $_{0.051}$ & & 0.134 & 0.092 & 0.043 & 0.068\\
    "Meeting of the two" & \textbf{0.660} $_{0.089}$ & \textbf{0.701} $_{0.085}$ & \textbf{0.740} $_{0.069}$ & \textbf{0.753} $_{0.069}$ & & 0.121 & 0.112 & 0.061 & 0.053\\
    \bottomrule
  \end{tabular}
    \caption{Inter-observer consistency of the measured temporal saliency maps. $C_{1-N}$ denotes sum of $N$ rounds used for computation.}
  \label{tab:temporal}
\end{table*}

\begin{figure*}[h!]
  \centering
  \includegraphics[width=\linewidth]{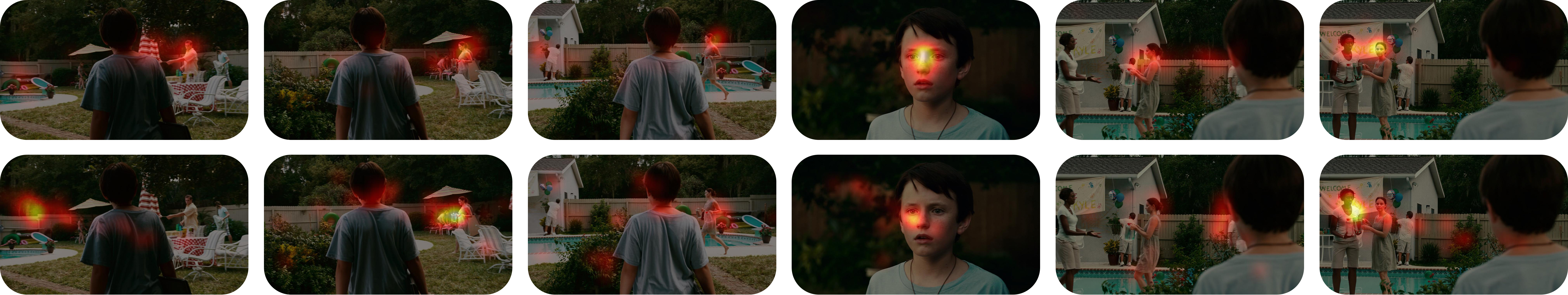}
  \caption{The comparison of spatial saliency maps. Top row in each pair -- eye-tracking results, bottom -- our results. Zoom is required.}
  \label{fig:spatial}
\end{figure*}

\section{Results and discussion}
The proposed interface allows measuring both temporal and spatial saliency at the same time, thus, we evaluate the accuracy of both these outputs.
\subsection{Temporal saliency results}
Considering that there are no ground truth temporal saliency data, we evaluate the output of the algorithm by \mbox{analyzing} the produced temporal saliency "maps" and estimating inter-observer consistency. The examples of obtained temporal saliency "maps" are illustrated in Fig. \ref{fig:temporal}. The demonstration of the videos with saliency scores encoded as a color-map is available online: \href{https://drive.google.com/open?id=1g7egplutXqOqSMtNl7N5NkO-4SkZTMxY}{[link]}. Figure \ref{fig:temporal} demonstrates three plots for each video which correspond to different averaging approaches: the sum of all clicks from all five video repeats ($C_{1-5}$); the sum of clicks only from the first round without repeating ($C_{1}$); and the \emph{weighted} sum of clicks from 5 rounds ($C_{1-5}^{(W)}=\sum_{n=1}^{5}C_n W_n$, where $W_n=\{1,0.8,0.6,0.4,0.2\}$). All the scores are normalized by a maximum number of clicks the frame can have.\\
 
Qualitative analysis shows that \emph{most of the peaks on the temporal saliency graph correspond to the semantically meaningful salient events on the video}. This is the main achievement of the proposed interface. It can also be seen that an intentionally taken monotonic video without salient events ("leaves in the wind") has relatively flat saliency graph without strongly pronounced peaks (which could be even flatter when the response statistics is larger). Apart from that, it may be seen that in the case of other videos, the output of the first round (red line) is very similar to the total output of all five rounds. This means that even when in the next rounds observers start exploring smaller, less salient details, they still return to the "main" events and follow a similar pattern of clicks as in the first round. Also, adding weights to the sum (thin black line) does not influence the results significantly, which again indicates the similarity of clicks from all the rounds. However, using $N$ rounds indeed allows to gather $N$ times more responses making the graph smoother and, as we show next, produces more consistent responses from each observer. \\
In order to estimate consistency between different groups of observers, we synthetically split observers into two groups of 15 people each. Then, we compute temporal saliency maps for each group independently and compare the results. The comparison is done using the Pearson Correlation Coefficient between the saliency maps from different groups, as well as performing the Kolmogorov-Smirnov test between two distributions and reporting the p-value. Results are averaged between 100 random splits (standard deviation is also reported for PCC). Table \ref{tab:temporal} shows that the correlation between responses from different observers is very high, up to 0.86. Increasing the number of rounds considered increases the correlation of responses significantly, with maximum values achieved when all five rounds are included.

\subsection{Spatial saliency results}
The spatial saliency maps produced by eye-tracking data versus our interface can be compared visually in Fig. \ref{fig:spatial}. (fixation points are blurred with a Gaussian of sigma equal to $1^{\circ}$ of visual angle (33 px)). As may be seen, the results are very similar, even though we did not use any special equipment and collected spatial data additionally to the main temporal output.  \\
Saliency maps are evaluated quantitatively using standard saliency metrics: Area under ROC Curve (AUC) \cite{judd2009learning}\cite{borji2013analysis} and Normalized Scanpath Saliency (NSS) \cite{peters2005components}. Table~\ref{tab:spatial} presents statistics of the scores computed per frame. Results demonstrate both good and poor performance, and differ significantly from video to video. Additionally, quality of spatial saliency can be assessed visually via the rendered videos with map overlay \href{https://drive.google.com/open?id=17Cjd1SwO0sqlkWbVm2F9WVd5zQHxkXOT}{[link]}, as well as the videos with both eye-tracking (blue dots) and our results (red dots) simultaneously \href{https://drive.google.com/open?id=1xRJ-U2O9zniXu0MQ7NTUvegPcoFP1XmI}{[link]}.

\begin{table}[t]
  \small
  \begin{tabular}{p{3.2cm}rr}
    \toprule
    & AUC (mean $_{std}$)& NSS (mean $_{std}$)\\
    \midrule
    "The underwater world"   &0.617 $_{0.108}$&0.73 $_{0.78}$\\
    "Cinematic scene"    &0.712 $_{0.119}$&1.59 $_{1.05}$\\
    "Leaves in the wind" &0.548 $_{0.055}$&0.18 $_{0.21}$\\
    "Basketball game"    &0.727 $_{0.114}$&1.52 $_{0.93}$\\
    "Diver suffocating"  &0.794 $_{0.113}$&2.66 $_{1.41}$\\
    "Meeting of the two" &0.625 $_{0.060}$&0.95 $_{0.43}$\\
  \bottomrule
  \end{tabular}
\caption{Comparison of the measured spatial saliency maps and gaze-fixations obtained using eye-tracker.}
  \label{tab:spatial}
\end{table}

\section{Conclusions}
In this work, we presented a novel mouse-contingent interface designed for measuring temporal and spatial video saliency. Temporal saliency is a novel concept which is studied incongruously less than it should in comparison to spatial saliency. Temporal video saliency allows identifying the important fragments of a video by assigning a saliency score to each frame. The analysis of the experimental study shows that the use of the proposed interface allows to accurately approximate the temporal saliency "map" as well as gaze-fixations of the observers at the same time.

\balance{} 

\bibliographystyle{SIGCHI-Reference-Format}
\bibliography{refs}

\end{document}